\documentclass[twocolumn,showpacs,preprintnumbers,amsmath,amssymb ]{revtex4}
\usepackage{graphicx}
\usepackage{float}

\begin{document}

\bibliographystyle{unsrt}

   \title{Vorticity, gyroscopic precession, and spin-curvature Force}
   \author{Wei Chieh Liang$^{}$}
     \email{d95943030@ntu.edu.tw}
   \author{Si Chen Lee$^{}$}
     \email{sclee@ntu.edu.tw}
     \affiliation{%
Graduate Institute of Electronics Engineering, 
National Taiwan University, Taipei, Taiwan
}%

\date{\today}

\begin{abstract}
In investigating the relationship between vorticity and gyroscopic precession, we calculate the vorticity
vector in Godel, Kerr, Lewis, Schwarzschild, and Minkowski metrics and find that the vorticity vector of
the specific observers is the angular velocity of the gyroscopic precession. Furthermore, when space-time
torsion is included, the vorticity and spin-curvature force change sign. This result is very similar to the
behavior of the positive and negative helicities of quantum spin in the Stern-Gerlach force. It implies that
the inclusion of torsion will lead to an analogous property of quantum spin even in classical treatment.
\end{abstract}

\maketitle
\section{Introduction}

The frame dragging effect (Lense-Thirring effect) was a prediction of general relativity that had been verified by the
Gravity Probe-B experiment \cite{everitt2011}. The effect can be measured by the precession of a gyroscope moving in a gravitational field with a rotating source (the Earth). The rotating source will drag the inertial frame (gyroscope)
and cause the gyroscopic precession with respect to the fixed stars \cite{gravitation,inertia}.

According to Smalley \cite{Smalley1985} who considered the Godel metric in Riemann-Cartan space-time, the appearance of torsion tensor induced by spinning fluid will not change the magnitude of the angular velocity of global rotation but results in the opposite sign. The spin flip of angular velocity was claimed to be consistent with the result of Tsoubelis \cite{Tsoubelies1984} who found that the spin density produced a global stationary space-time outside a static but spin-polarized cylinder space-time. In that case, the spin is responsible for the presence of t-$\phi$ cross terms (t,r,$\theta$,$\phi$ are coordinates) in the metric field outside. It reflects the dragging of inertia. Thus, spin has a status equal to that of orbital angular momentum.

After further consideration, it is found that the so-called spin flip is an ambiguous concept. In their model, the torsion is caused by fluid with intrinsic but classical spin. In normal Einstein-Cartan theory, torsion is associated with quantum mechanical spin. It is not clear, what the true meaning of global rotation with an opposite sign is. 

The global angular velocity of the Godel universe is actually the vorticity vector\cite{Smalley1985}.Following this idea, the
vorticity vector of Kerr, Lewis, Schwarzschild, and Minkowski (rotating coordinate) metrics in Riemann and Riemann-Cartan space-times was investigated to find out the deeper physical meaning of vorticity and torsion. First, the basic knowledge of relativistic gyroscopic precession, Riemann-Cartan space-time, and Einstein-Cartan-Sciama-Kibble (ECSK) theory is reviewed and the meaning of vorticity is discussed.  The vorticity vectors of each metric
by selecting a specific observer in Riemann and Riemann-Cartan space-times are calculated; then the spin-curvature force from the Mathisson-Papapetrou equation of the Godel metric is analyzed. Finally, the conclusions are given.
\section{Preliminaries}
\subsection{Gyroscopic precession}
The spin angular momentum of a body will be affected by the curvature of the space-time, so a spinning gyroscope orbiting a massive body will undergo precession of its spin vector S: $dS/dt = \Omega  \times S, \Omega$ is the angular velocity of precession. There are three relativistic contributions to this precession \cite{everitt2011,gravitation,inertia}.
\begin{figure}[H]
\center
\includegraphics[width=8cm]{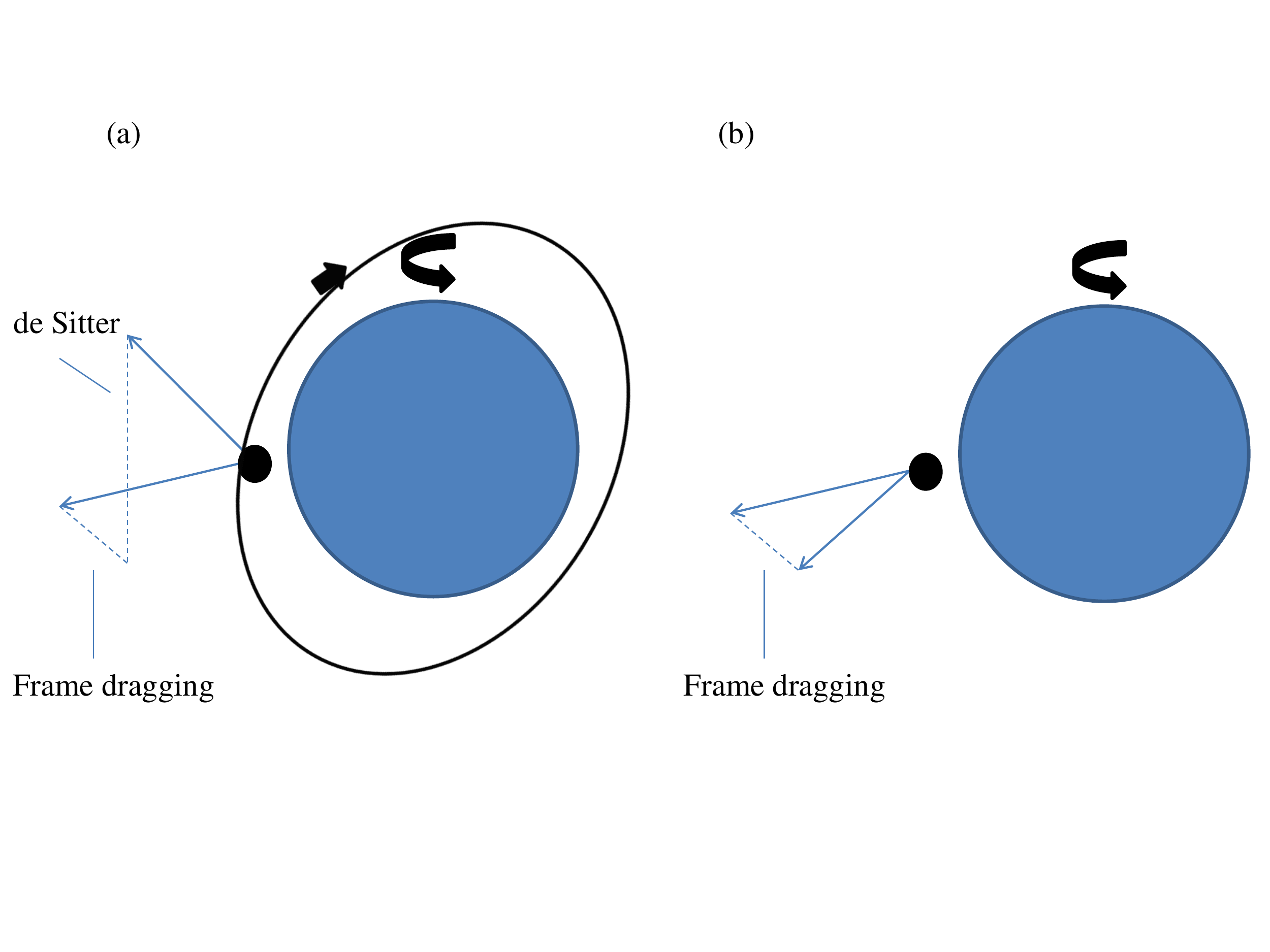}\\
  \caption{(a) When the gyroscope moves around a rotating mass, it will undergo both de Sitter and frame dragging precessions. (b) If the gyroscope is at a static point near the rotating mass, it only feels the precession due to the dragging of the space-time.}
\end{figure}
1.	The Thomas precession, arising along a curved trajectory as an additional rotation accompanying the composition of nonaligned Lorentz boosts, which is a special relativistic effect.
2.	The de Sitter precession, associated with the gyroscopic transportation in curved space-time generates a precession with respect to a fixed reference.
3.	The frame dragging precession, arising from the rotation of the central body which causes the inertial frames to be dragged along with respect to the fixed stars. It is of purely geometrical origin and is independent of the orbital elements of the gyroscope as shown in Fig. 1(a) and (b)
\subsection{The Riemann-Cartan space-time (U4 geometry) and ECSK theory}
The most natural modification of general relativity in Riemannian geometry (V4) is ECSK theory. This geometry is no longer Riemannian but Riemann-Cartan space-time with a nonsymmetric and metric compatible connection. It brings up a new
geometrical property of space-time, which is the torsion tensor \cite{Hehl1976}.
The definition of torsion tensor $T_{\alpha \beta }^\mu$ is
\begin{align}
{T_{\alpha \beta }}^\mu  = \frac{1}
{2}({\Gamma _{\alpha \beta }}^\mu  - {\Gamma _{\beta \alpha }}^\mu ),
\end{align}
where $\Gamma_{\alpha \beta }^\mu$ are the connection coefficients.
In Einstein-Cartan theory, the intrinsic angular momentum (particle spin) is the source of space-time torsion \cite{Hehl1976}. The connection of U4 $\Gamma _{\alpha \beta \mu }$ is different from the Levi-Civita connection in V4 ${\tilde \Gamma _{\alpha \beta \mu }}$.

\begin{align}
{\Gamma _{\alpha \beta \mu }} = {\tilde \Gamma _{\alpha \beta \mu }} - {K_{\alpha \beta \mu }},
\end{align}
the second term is the contortion tensor which is given by
\begin{align}
{K_{\alpha \beta \mu }} =  - {T_{\alpha \beta \mu }} + {T_{\beta \mu \alpha }} - {T_{\mu \alpha \beta }}.
\end{align}
In anholonomy basis, the torsion tensor becomes
 \begin{align}
{T_{\hat a\hat b}}^{\hat c} = \frac{1}
{2}({\Gamma _{\hat a\hat b}}^{\hat c} - {\Gamma _{\hat b\hat a}}^{\hat c}) + {f_{\hat a\hat b}}^{\hat c}
\end{align}
\begin{align}
{f_{\hat a\hat b}}^{\hat c} = \frac{1}
{2}{e_{\hat a}}^\mu {e_{\hat b}}^\nu ({\partial _\mu }{e^{\hat c}}_\nu  - {\partial _\nu }{e^{\hat c}}_\mu ),
\end{align} 
where ${f_{\hat a\hat b}}^{\hat c}$ is the anholonomity and ${e_{\hat a}}^\mu$ is tetrad frame.
The torsion tensor can be decomposed into three irreducible parts \cite{S.Capozziello2001}
\begin{align}
{T_{\mu \nu }}^\lambda  = {\tilde T_{\mu \nu }}^\lambda  + \frac{2}
{3}\delta _{{\text{[}}\nu }^\lambda {T_{\mu {\text{]}}}} + \frac{1}
{3}{\varepsilon _{\mu \nu \beta }}^\lambda {\hat T^\beta },
\end{align} 
where
\begin{align}
{T_\mu } = {T_{\mu \lambda }}^\lambda
\end{align}
is the trace part of torsion tensor(V). It has no relation to the spin density but will preserve the minimal coupling and gauge invariance of the electromagnetic field in ECSK theory \cite{S.Hojman1978}.
\begin{align}
{\hat T^\alpha } = \frac{1}
{{3!}}{\varepsilon ^\alpha }_{\mu \nu \lambda }{T^{\mu \nu \lambda }}
\end{align}
is the axial torsion vector. The last term of Eq.(6) is the totally antisymmetric part of torsion tensor(A), which is believed to couple with Dirac particles \cite{S.Capozziello2001}, and ${\tilde T_{\mu \nu }}^\lambda $ is the rest of the traceless part but not totally antisymmetric(TL).
One defined the modified torsion tensor
\begin{align}
{Q_{\lambda \mu }}^\nu  = {T_{\lambda \mu }}^\nu  + \delta _\lambda ^\nu {T_\mu } - \delta _\mu ^\nu {T_\nu }.
\end{align}
In V4, the Levi-Civita connection is determined by the metric compatible and torsion-free condition. In U4, the torsion-free condition is eliminated, so there is no unique solution of the metric compatible connection except some proper constrained conditions are chosen. There are many ways to select a torsion tensor, for example, the teleparallel gauge \cite{K.Hayashi1979, Hammond2002, Mao2007}, the torsion arising from polarized spin \cite{Harald1990, Obukhov1987}, the spherical symmetric condition \cite{Yasskin1979, BAEKLER1981}, and the specific torsion from three irreducible parts \cite{Hammond1996, S.Capozziello2001}.     

In ECSK theory, the Einstein-Cartan-Sciama-Kibble equations are given by
\begin{align}
\begin{gathered}
  {G_{\mu \nu }} = k{\theta _{\mu \nu }} \hfill \\
  {Q_{\lambda \mu \nu }} = k{S_{\lambda \mu \nu }} \hfill \\ 
\end{gathered},
\end{align}
where $G_{\mu\nu}$ is the Einstein tensor with respect to the U4 connection.
\begin{align}
{\theta ^{\mu \nu }} = {T^{\mu \nu }} + {\nabla ^*}_\alpha {\text{(}}{{\text{S}}^{\mu \nu \alpha }} - {S^{\nu \alpha \mu }} + {S^{\alpha \mu \nu }}{\text{)}}
\end{align}
is the energy momentum tensor $T_{\mu\nu}$ plus extra spin contributions where $S^{\alpha\mu\nu}$ is the spin density tensor and ${\nabla ^*}_\alpha  = {\nabla _\alpha } + 2{T_\alpha }$, where ${\nabla _\alpha }$ is a covariant derivative with respect to the U4 connection.

In the framework of the early ECSK theory, they introduced a Weyssenhoff fluid to model a spinning perfect fluid as a source for the torsion \cite{F.W.Hehl1974}. Ray and Smalley provided a self-consistent variational principle for a Weyssenhoff spinning fluid in V4 and U4 \cite{R.Ray1983, Krisch1993}, where the spinning fluid is given by
\begin{align}
\begin{gathered}
  {T_{\mu \nu }} = {T_{\mu \nu }}(fluid) + {T_{\mu \nu }}(spin) \hfill \\
  {T_{\mu \nu }}(fluid) = \rho (1 + \varepsilon  + p/\rho ){U_\mu }{U_\nu } + {g_{\mu \nu }}p\ \hfill \\
{T_{\mu \nu }}(spin) = {U_{(\mu }}{S_{\nu )\sigma }}{{\dot U}^\sigma } + {\nabla _\alpha }[{U_{(\nu }}{S_{\mu )\alpha }}] \\ 
  - {{\tilde \omega }_{\sigma (\mu }}{S_{\nu )}} + {U_{(\mu }}{S_{\nu )\sigma }}{{\tilde \omega }^{\mu \alpha }}{U_\alpha } \\ 
\end{gathered}
\end{align}
where $\rho$ is the matter density, $\varepsilon$ is the internal energy density, p is the pressure, U is four velocity, ${\dot U^\sigma } = {U^\alpha }{\nabla _\alpha }{U^\sigma }$, and ${\tilde \omega _{\mu \nu }}$ is the angular velocity associated with the spin density which can be defined as 
\begin{align}
{\tilde \omega _{\mu \nu }} = {U^\sigma }{\nabla _\sigma }{e^{\hat a}}_\mu  \cdot {e_{\hat a}}_\nu.
\end{align}
It is equal to the vorticity vector when the acceleration is zero.
The spin tensor is defined as
\begin{align}
{S_{\mu \nu }} = \rho \kappa({e^{\hat 1}}_\mu {e^{\hat 2}}_\nu  - {e^{\hat 1}}_\nu {e^{\hat 2}}_\mu ),
\end{align}
where $\kappa$ is a scalar function and the relation between the spin density and spin tensor is
\begin{align}
{S_{\mu \nu }}^\sigma  = {S_{\mu \nu }}{U^\sigma }.
\end{align}
The model of Godel cosmology in Riemann-Cartan space-time was based on this source by Smalley \cite{Smalley1985}.

In this paper, the tetrad filed is denoted by $e^{\hat a}_\mu$  . The Greek alphabet ($\mu$, $\nu$, $\rho$, ...= 0, 1, 2, 3) are used to denote tensor indices, The Latin alphabet with hat (${\hat a}$,${\hat b}$, ${\hat c}$, ...=0, 1, 2, 3) will be used to denote frame indices. The Minkowski space-time metric tensor raises and lowers tetrad indices and is fixed by ${\eta _{ab}} = {e_{\hat a\mu }}{e_{\hat b\nu }}{g^{\mu \nu }} = ( -  +  +  + )$.
\subsection{Vorticity}
In relativistic hydrodynamics, the gradient of the velocity field u is decomposed in the following way \cite{gravitation,inertia}.
\begin{align}
{\nabla _\mu }{u_\nu } = \frac{1}{3}\theta {h_{\mu \nu }} + {\sigma _{\mu \nu }} + {\omega _{\mu \nu }} + {u_\mu }{a_\nu },
\end{align}
where ${h_{\alpha \beta }} = {g_{\alpha \beta }} + {u_\alpha }{u_\beta }$ is the space-projection tensor.
\begin{align}
\begin{gathered}
  \theta  = {\nabla _\alpha }{u^\alpha } \hfill \\
  {\sigma _{\mu \nu }} = h_\mu ^\alpha h_\nu ^\beta {\text{(}}{\nabla _{(\alpha }}{u_{\beta )}} - \frac{1}
{3}\theta {h_{\alpha \beta }}{\text{)}} \hfill \\
  {a_\nu } = {u^\alpha }{\nabla _\alpha }{u_\nu } \hfill, \\ 
\end{gathered}
\end{align}
which are the expansion scalar, shear tensor, and acceleration vector, respectively.
The vorticity tensor is defined as
\begin{align}
{\omega _{\mu \nu }} = h_\mu ^\alpha h_\nu ^\beta {\nabla _{[\alpha }}{u_{\beta ]}}.
\end{align}
${\nabla _\alpha }$ is the covariant derivative in U4, but all the equations have the same form in V4.
The vorticity vector in V4 is given by
\begin{align}
{\omega ^\alpha } = \frac{1}{2}{\varepsilon ^{\alpha \beta \mu \nu }}{u_\beta }{\omega _{\mu \nu }} = \frac{1}{2}{\varepsilon ^{\alpha \beta \mu \nu }}{u_\beta }{u_{\mu ;\nu }} = \frac{1}{2}{\varepsilon ^{\alpha \beta \mu \nu }}{u_\beta }{u_{\mu ,\nu }}
\end{align}
where the semicolon and comma are the covariant and partial derivatives. It has the following identity\cite{Ehlers1993} :
\begin{align}
{\tilde \nabla _\alpha }{\omega ^\alpha } = 2{a_\alpha }{\omega ^\alpha },
\end{align}
where ${\tilde\nabla}$ is the covariant derivative in V4.

Vorticity geometrically measures the twisting of the congruence \cite{B.R.Iyer1993}. The meaning of the vorticity vector is
the rotation of the connecting vector as it relates to the Fermi-Walker frame. If a space-time with a timelike Killing vector $\xi$ is considered, the four velocity of the observer is ${u^\alpha } = {\xi ^\alpha }{\rm{/}}\sqrt { - {\xi ^\beta }{\xi _\beta }}$. Each observer arranges his spatial basis vectors so that they connect to the same neighboring observers for all time t, which means the spatial basis is Lie-dragged along $u$. Lie-dragging locks the spatial frame to connecting vectors. That means the vorticity $\omega$ is equal to the rotation relative to the Fermi-Walker frame which is physically realized by the system of gyroscopes. Therefore, the angular velocity of the gyroscopic precession relative to the reference frame is given by $-\omega$ \cite{gravitation,B.R.Iyer1993}.

From Eq.(19), we can see there is no difference between the vorticity vectors in V4 and flat space-time. It seems that V4 space-time cannot reflect the geometric property of the vorticity vector unless the torsion tensor exists.
The vorticity vector with torsion in U4 is
\begin{align}
\begin{gathered}
  {\omega ^\alpha }_{(T)} = \frac{1}
{2}{\varepsilon ^{\alpha \beta \mu \nu }}{u_\beta }{\nabla _\nu }{u_\mu } \hfill \\
   = \frac{1}
{2}{\varepsilon ^{\alpha \beta \mu \nu }}{u_\beta }[{\partial _\nu }{u_\mu } - ({{\tilde \Gamma }_{\nu \mu }}^\gamma  - {K_{\nu \mu }}^\gamma ){u_\gamma }] \hfill \\
   = \frac{1}
{2}{\varepsilon ^{\alpha \beta \mu \nu }}{u_\beta }({u_{\mu ;\nu }} + {K_{\nu \mu }}^\gamma {u_\gamma }) \hfill, \\ 
\end{gathered}
\end{align}
where ${\Gamma _{\nu \mu }}^\gamma$ and ${\tilde \Gamma _{\nu \mu }}^\gamma$ are the connections of U4 and V4.
\section{Calculation of Vorticity Vector and Spin-Curvature Force}
\subsection{Godel metric}
The result derived by Smalley \cite{Smalley1985} could also be calculated alternately using the simple method which is to calculate Eq.(21) directly.
The Godel metric is \cite{Godel1949}
\begin{align}
d{s^2} = {a^2}[ - {(dt + {e^x}dy)^2} + d{x^2} + \frac{1}{2}{e^{2x}}d{y^2} + d{z^2}],
\end{align}
where $a$ is constant. 
The tetrad is chosen as
\begin{align}
\begin{gathered}
  {e_{\hat t}} = \frac{1}
{a}{\partial _t},{e_{\hat x}} = \frac{{\text{1}}}
{a}{\partial _x},{e_{\hat y}} = \frac{{ - \sqrt 2 }}
{a}({\partial _t} - {e^{-x}}{\partial _y}),{e_{\hat z}} = \frac{1}
{a}{\partial _z} \hfill \\
  {e^{\hat t}} = adt + a{e^x}dtdy,{e^{\hat x}} = adx,{e^{\hat y}} = (\frac{{a{e^x}}}
{{\sqrt 2 }})dy,{e^{\hat z}} = adz. \hfill \\
\end{gathered}
\end{align}
The four velocity is defined as ${u^\beta } = e_{\hat t}^\beta$, so 
\begin{align}
{u^\beta } = (  \frac{1}{a},0,0,0).
\end{align}
The vorticity vector in V4 is given by Eq.(19); the result is
\begin{align}
{\omega ^\mu } = (0,0,0, - \frac{1}
{2}\frac{{\sqrt 2 }}
{{{a^2}}}).
\end{align}
Now, we consider the case in U4. The treatment by Smalley \cite{Smalley1985} to decide the torsion tensor was based on two assumptions. First, the torsion tensor in anholonomy basis is proportional to the anholonomity
\begin{align}
{T_{\hat a\hat b}}^{\hat c} = \frac{1}
{2}({\Gamma _{\hat a\hat b}}^{\hat c} - {\Gamma _{\hat b\hat a}}^{\hat c}) + {f_{\hat a\hat b}}^{\hat c} = C \cdot {f_{\hat a\hat b}}^{\hat c},
\end{align}
where $C$ is constant. This assumption is an extension of teleparallel gauge. There are still other similar choices, for example, the torsion where $C$ is confined between 0-1 and the connection ${\Gamma}=(1-C){\tilde \Gamma}-C\cdot K$ describes the change between general relativity (GR)($C = 0$) and teleparallel theory ($C=1$)\cite{Mao2007}, the torsion components composed of the partial derivative of tetrad field which is more general than Eq.(26)\cite{VULCANOV2000}. 
Second, the trace part of the torsion tensor, Eq.(7), will vanish because the spin fluid will not couple to that part.
The anholonomity of the tetrad is 
\begin{align}
\begin{gathered}
  {f_{\hat x\hat y}}^{\hat t} =  - {f_{\hat y\hat x}}^{\hat t} = \frac{{\sqrt 2 }}
{{2a}} \hfill \\
  {f_{\hat x\hat y}}^{\hat y} =  - {f_{\hat y\hat x}}^{\hat y} = \frac{1}
{{2a}} \hfill \\ 
\end{gathered}.
\end{align}
According to his model, there is a nonvanishing component of spin tensor ${S_{\hat x\hat y}}$; thus, $C \cdot {S_{\hat x\hat y}}^{\hat t} = C \cdot {S_{\hat x\hat y}}{u^{\hat t}} = {T_{\hat x\hat y}}^{\hat t} \ne 0$, but ${T_{\hat x\hat y}}^{\hat y} = 0$ due to ${u^{\hat y}} = 0$. In order to make it consistent, he set ${T_{\hat x\hat y}}^{\hat y} = 0$, although $C \cdot {f_{\hat x\hat y}}^{\hat y}$ is not zero. This assumption will zero not only the trace part of the torsion but also the traceless part of torsion which is not totally antisymmetrical. The only nonvanishing torsion tensor now is given by
\begin{align}
{T_{\hat x\hat y}}^{\hat t} = C{f_{\hat x\hat y}}^{\hat t} =  - {T_{\hat y\hat x}}^{\hat t} = C\frac{{\sqrt 2 }}
{{2a}}.
\end{align}
It is transformed to the holonomic basis
\begin{align}
{T_{xy}}^t =  - {T_{yx}}^t = \frac{{  C{e^x}}}
{2}.
\end{align}
Using Eq.(21), the vorticity vector with torsion is
\begin{align}
{\omega ^\mu }_{(T)} = (0,0,0,\frac{1}
{2}\frac{{( - 1 + C)\sqrt 2 }}
{{{a^2}}}).
\end{align}
When $C= 0$ and torsion=0, Eq.(30) is identical to the vorticity vector in V4.
When $C=2$,
\begin{align}
{\omega ^\mu }_{(T)} =  - {\omega ^\mu },
\end{align}
which is the same as the result of Smalley \cite{Smalley1985}.
It should be mentioned that when $C=2$, the spinning fluid model of Ray and Smalley is consistent with the Godel cosmology in U4. This choice of torsion tensor, Eq.(26), has many advantages. It contains the teleparallel case which is a reformulation of GR in terms of the tetrad field. By selecting the proper constant, it will preserve the physics of GR in V4 and bring up a new property in U4. Furthermore,  the information of different sources of torsion can be obtained through its three irreducible parts, which will be demonstrated in the following sections. 

\subsection{Kerr metric}
In Boyer-Linquist coordinates, the Kerr metric takes the form
\begin{align}
\begin{gathered}
  d{s^2} =  - (1 - \frac{{2mr}}
{\Sigma })d{t^2} + \frac{\Sigma }
{\Delta }d{r^2} + \Sigma d{\theta ^2} \hfill \\
   - \frac{{4mra{{\sin }^2}\theta }}
{\Sigma }d\phi dt + ({r^2} + {a^2} + \frac{{2mr{a^2}{{\sin }^2}\theta }}
{\Sigma }){\sin ^2}\theta d{\phi ^2} \hfill \\
  \Sigma  = {r^2} + {a^2}{\cos ^2}\theta \begin{array}{*{20}{c}}
   {} & {\Delta  = {r^2} - 2mr + {a^2}}  \\
\end{array}  \hfill \\ 
\end{gathered}
\end{align}
The local static frame \cite{Semerak1993} is given by
\begin{align}
\begin{gathered}
  {e_{\hat t}} = \sqrt {\frac{\Sigma }
{{\Sigma  - 2mr}}} {\partial _t},{e_{\hat r}} = \sqrt {\frac{\Delta }
{\Sigma }} {\partial _r},{e_{\hat \theta }} = \frac{1}
{{\sqrt \Sigma  }}{\partial _\theta }, \hfill \\
  {e_{\hat \phi }} = \frac{{ - 2mar\sin \theta }}
{{\sqrt {\Delta \Sigma (\Sigma  - 2mr)} }}{\partial _t} + \frac{{\sqrt {\Sigma  - 2mr} }}
{{\sin \theta \sqrt {\Delta \Sigma } }}{\partial _\phi } \hfill \\
  {e^{\hat t}} = \sqrt {\frac{\Sigma }
{{\Sigma  - 2mr}}} (\frac{{\Sigma  - 2mr}}
{\Sigma }dt + \frac{{2mar{{\sin }^2}\theta }}
{\Sigma }d\phi ), \hfill \\
  {e^{\hat r}} = \sqrt {\frac{\Sigma }
{\Delta }} dr,{e^{\hat \theta }} = \sqrt \Sigma  d\theta ,{e^{\hat \phi }} = \frac{{\Sigma \Delta \sin \theta }}
{{\sqrt {\Delta \Sigma (\Sigma  - 2mr)} }}d\phi  \hfill \\ 
\end{gathered}.
\end{align}
The four velocity is
\begin{align}
{u^\alpha } = (\sqrt {\frac{\Sigma }
{{\Sigma  - 2mr}}} ,0,0,0),
\end{align}
then, the nonzero components of vorticity vector in V4 are
\begin{align}
\begin{gathered}
  {\omega ^r} =  - \frac{{2mra\cos \theta ({r^2} - 2mr + {a^2})}}
{{{{({r^2} + {a^2}{{\cos }^2}\theta )}^2}({r^2} - 2mr + {a^2}{{\cos }^2}\theta )}} \hfill \\
  {\omega ^\theta } =  - \frac{{ma\sin \theta ({r^2} - {a^2}{{\cos }^2}\theta )}}
{{{{({r^2} + {a^2}{{\cos }^2}\theta )}^2}({r^2} - 2mr + {a^2}{{\cos }^2}\theta )}} \hfill \\ 
\end{gathered}.
\end{align}
In U4, now we will slightly modify the torsion used in Godel metric to the whole traceless part of the torsion; thus,
\begin{align}
{T_{\mu \nu }}{^\lambda _{(A + TL)}} = {T_{\mu \nu }}^\lambda  - \frac{2}
{3}\delta _{{\text{[}}\nu }^\lambda {T_{\mu {\text{]}}}}.
\end{align}
The vorticity vector with torsion turns to
\begin{align}
\begin{gathered}
  {\omega ^r}_{(T)} =  \frac{{(C - 1)2mra\cos \theta ({r^2} - 2mr + {a^2})}}
{{{{({r^2} + {a^2}{{\cos }^2}\theta )}^2}({r^2} - 2mr + {a^2}{{\cos }^2}\theta )}} \hfill \\
  {\omega ^\theta }_{(T)} = \frac{{(C - 1)ma\sin \theta ({r^2} - {a^2}{{\cos }^2}\theta )}}
{{{{({r^2} + {a^2}{{\cos }^2}\theta )}^2}({r^2} - 2mr + {a^2}{{\cos }^2}\theta )}} \hfill \\ 
\end{gathered}.
\end{align}
When $C= 0$ and torsion=0, we get the vorticity vector in V4.
When $C=2$, ${\omega ^\alpha }_{(T)} =  - {\omega ^\alpha }$.

The next problem that needs to be considered is the meaning of the vorticity vector of the static observer in Kerr space-time. Considering the case at the equator $\theta=\pi/2$,
\begin{align}
{\omega ^\alpha } = (0,0,\frac{{ - ma}}
{{{r^3}(r - 2m)}},0),
\end{align}
and the scalar vorticity
\begin{align}
\Omega  = {({\omega ^\alpha }{\omega _\alpha })^{1/2}} = \frac{{ma}}
{{{r^2}(r - 2m)}} = \frac{{ma}}
{{{r^3}}}{(1 - \frac{{2m}}
{r})^{ - 1}}
\end{align}
is the precession of the gyroscope due to frame dragging at the equator, which is the same as the result of Iyer and Vishveshawara calculated by the Frenet-Serret method \cite{B.R.Iyer1993}. This is because we choose static observers(see Fig. 1) which are proportional to the time-like Killing vector,whose base’s frame vectors are lie dragged along $u$,
\begin{align}
\begin{gathered}
  P(u){L_u}{e_i} = 0 \hfill \\
  P(u) = {h^\alpha }_\beta  \hfill. \\ 
\end{gathered}
\end{align}
It locks the spatial frame to the connecting vector of the Killing congruence, which is always pointed to the same fixed stars \cite{D.Bini2005,F.D.Felice2010}.
\subsection{Lewis metric}
The general stationary axisymmetric metric considered is the standard form of the rotating metric \cite{Islam1985}.
\begin{align}
d{s^2} =  - fd{t^2} - 2kdtd\phi  + ld{\phi ^2} + {e^b}(d{r^2} + d{z^2})
\end{align}
with coordinates ($t,r,z,\phi$).
The metric potentials are only functions of r.
The local static frame is given by \cite{Krisch1987}
\begin{align}
\begin{gathered}
  {e_{\hat t}} = 1/\sqrt {f(r)} {\partial _t},{e_{\hat r}} = {e^{ - b(r)/2}}{\partial _r},{e_{\hat z}} = {e^{ - b(r)/2}}{\partial _z}, \hfill \\
  {e_{\hat \phi }} = \frac{1}
{{\sqrt {f(r)l(r) + k{{(r)}^2}} }}(\frac{{ - k(r)}}
{{\sqrt {f(r)} }}{\partial _t} + \sqrt {f(r)} {\partial _\phi }) \hfill \\
  {e^{\hat t}} = \sqrt {f(r)} dt + \frac{{k(r)}}
{{\sqrt {f(r)} }}d\phi ,{e^{\hat r}} = {e^{b(r)/2}}dr, \hfill \\
  {e^{\hat z}} = {e^{b(r)/2}}dz,{e^{\hat \phi }} = \frac{{\sqrt {f(r)l(r) + k{{(r)}^2}} }}
{{\sqrt {f(r)} }}d\phi  \hfill. \\ 
\end{gathered}
\end{align}
The four velocity
\begin{align}
{u^\alpha } = (\frac{1}
{{\sqrt {f(r)} }},0,0,0).
\end{align}
The vorticity vector in V4 is
\begin{align}
{\omega ^\alpha } = (0,0,\frac{1}
{2}\frac{{[f(r)k'(r) - k(r)f'(r)]}}
{{{e^{b(r)}}f(r)\sqrt {f(r)l(r) + k{{(r)}^2}} }},0);
\end{align}
and the vorticity vector in U4 with traceless part of torsion becomes
\begin{align}
{\omega ^\mu }_{(T)} = (0,0,\frac{1}
{2}\frac{{(1 - C)[f(r)k'(r) - k(r)f'(r)]}}
{{{e^{b(r)}}f(r)\sqrt {f(r)l(r) + k{{(r)}^2}} }},0).
\end{align}
When $C= 0$ and torsion=0, Eq.(45) is the vorticity vector in V4.
When $C=2$, ${\omega ^\alpha }_{(T)} =  - {\omega ^\alpha }$.
We still want to know the meaning of the vorticity vector belonging to this observer.
The scalar vorticity
\begin{align}
\Omega  ={({\omega _\mu }{\omega ^\mu })^{1/2}}= \frac{{{e^{ - b(r)/2}}\left| {k(r)f'(r) - f(r)k'(r)} \right|}}
{{2f(r)\sqrt {f(r)l(r) + k{{(r)}^2}} }}
\end{align}
is the precession angular velocity of the gyroscope due to frame dragging in Lewis space-time, which is the same as the result of Herrera \cite{L.Herrera2000}.
\subsection{Schwarzschild metric}
The well-known Schwarzschild metric is given by
\begin{align}
d{s^2} =  - (1 - \frac{{2m}}
{r})d{t^2} + \frac{1}
{{1 - \frac{{2m}}
{r}}}d{r^2} + {r^2}d{\theta ^2} + {r^2}{\sin ^2}\theta d{\phi ^2}.
\end{align}
The four velocity of the static observer in the Schwarzschild metric is hypersurface orthogonal, which means the vorticity is zero. Therefore, we have to choose such an observer who has a nonzero and physically meaningful vorticity. According to de Felice and Ussegllo-Tomasset \cite{F.deFelice1992}, the tetrad was chosen as
\begin{align}
\begin{gathered}
  {e_{\hat t}} = \frac{1}
{{r\sqrt {\Lambda  - {w^2}{{\sin }^2}\theta } }}({\partial _t} + w{\partial _\phi }),{e_{\hat r}} = \Xi {\partial _r},{e_{\hat \theta }} = 1/r, \hfill \\
  {e_{\hat \phi }} = \frac{1}{{\sqrt {\Lambda  - {w^2}{{\sin }^2}\theta } }}(\frac{{w\sin \theta }}{\Xi }{\partial _t} 
  +\frac{\Xi }{{{r^2}\sin \theta }}{\partial _\phi }) \hfill; \\
  {e^{\hat t}} = \frac{1}{{\sqrt {\Lambda - {w^2}{{\sin }^2}\theta } }}(\frac{{r - 2m}}{{{r^2}}}dt
   +  - wr{\sin ^2}\theta d\phi ), \hfill \\
  {e^{\hat r}} =  - \frac{{r\Xi }}{{ - r + 2m}}dr,{e^{\hat \theta }} = rd\theta , \hfill \\
  {e^{\hat \phi }} = \frac{{\sin \theta }}
{{\sqrt {\Lambda - {w^2}{{\sin }^2}\theta } }}(\frac{{( - r + 2m)w}}
{{r\Xi }}dt + \Xi d\phi ) \hfill. \\ 
\end{gathered}
\end{align}
where $\Lambda  = \frac{{1 - 2m/r}}
{{{r^2}}}$, $\Xi  = \sqrt {1 - \frac{{2m}}
{r}}$, and $w$ is the orbital angular velocity.
The four velocicity is
\begin{align}
{u^\alpha } = (\frac{1}
{{r\sqrt {\Lambda  - {w^2}{{\sin }^2}\theta } }},0,0,\frac{w}
{{r\sqrt {\Lambda  - {w^2}{{\sin }^2}\theta } }}).
\end{align}
Then, the nonzero components of vorticity vector in V4 are
\begin{align}
\begin{gathered}
  {\omega ^r} = \frac{{( - r + 2m)w\cos \theta }}
{{ - r + 2m + {w^2}{r^3}{{\sin }^2}\theta }} \hfill \\
  {\omega ^\theta } =  - \frac{{( - r + 3m)w\sin \theta }}
{{r( - r + 2m + {w^2}{r^3}{{\sin }^2}\theta )}} \hfill. \\ 
\end{gathered}
\end{align}
Considering the case at the equator $\theta=\pi/2$
\begin{align}
{\omega ^\mu } = (0,0,\frac{{w( - r + 3m)}}
{{r( - r + 2m + {w^2}{r^3})}},0),
\end{align}
\begin{align}
\Omega  = {({\omega _\mu }{\omega ^\mu })^{1/2}} = \frac{{w( - r + 3m)}}
{{ - r + 2m + {w^2}{r^3}}},
\end{align}
which is the gyroscopic precession at the equator \cite{B.R.Iyer1993}.
Along a geodesic trajectory, the acceleration is zero, which leads to the Keplerian frequency
${w^2} = m/{r^3}$, thus $\Omega  = w$, which is the same as the result of Rindler and Perlick \cite{W.Rindler1990}. 
It brings out the de Sitter precession due to the space-time curvature.

In U4, the vorticity vector with the traceless torsion is
\begin{align}
\begin{gathered}
  {\omega ^r}_{(T)} =  - \frac{{( - 1 + C)( - r + 2m)w\cos \theta }}
{{ - r + 2m + {w^2}{r^3}{{\sin }^2}\theta }} \hfill \\
  {\omega ^\theta }_{(T)} = \frac{{( - 1 + C)( - r + 3m)w\sin \theta }}
{{r( - r + 2m + {w^2}{r^3}{{\sin }^2}\theta )}} \hfill. \\ 
\end{gathered}
\end{align}
When $C= 0$ and torsion=0, the vorticity vector in V4 is obtained. When $C=2$, the vorticity changes to the opposite sign.
\subsection{Minkowski metric in rotating coordinates}
The transform $d\phi  \to d\phi  + \omega dt$ brings the Minkowski metric into the rotating form with coordinates ($t,r,\phi,z$) \cite{W.Rindler1990}.
\begin{align}
\begin{gathered}
  d{s^2} =  - (1 - {w^2}{r^2}){[dt -w \zeta (r)d\phi ]^2} + d{r^2} \hfill \\
   +\zeta (r)  d{\phi ^2} + d{z^2} \hfill. \\ 
\end{gathered} 
\end{align}
where $\zeta (r) = \frac{{{r^2}}}{{1 - {w^2}{r^2}}}$. 
The frame of the static observer is
\begin{align}
\begin{gathered}
  {e_{\hat t}} = \frac{1}
{{\sqrt {1 - {w^2}{r^2}} }}{\partial _t},{e_{\hat r}} = {\partial _r}, \hfill \\
  {e_{\hat \phi }} = \frac{{wr}}
{{\sqrt {1 - {w^2}{r^2}} }}{\partial _t} +  \frac{\sqrt {1 - {w^2}{r^2}}}
{r}{\partial _\phi },{e_{\hat z}} = {\partial _z} \hfill, \\
  {e^{\hat t}} = \frac{{1 - {w^2}{r^2}}}
{{\sqrt {1 - {w^2}{r^2}} }}dt - \frac{{w{r^2}}}
{{\sqrt {1 - {w^2}{r^2}} }}d\phi , \hfill \\
  {e^{\hat r}} = dr,{e^{\hat \phi }} = \frac{r}
{{\sqrt {1 - {w^2}{r^2}} }}d\phi ,{e^{\hat z}} = dz \hfill. \\ 
\end{gathered}
\end{align}
The four velocity
\begin{align}
{u^\alpha } = (\frac{1}
{{\sqrt {1 - {w^2}{r^2}} }},0,0,0).
\end{align}
The vorticity vector in V4 is
\begin{align}
{\omega ^\alpha } = (0,0,0,\frac{w}
{{1 - {w^2}{r^2}}}).
\end{align}
The vorticity scalar
\begin{align}
\Omega  = {({\omega _\nu }{\omega ^\nu })^{1/2}} = \frac{w}
{{1 - {w^2}{r^2}}},
\end{align}
which is the gyroscopic precession in Minkowski space-time, or the Thomas precession angular velocity \cite{W.Rindler1990}.
Again, in U4, the vorticity vector with the traceless torsion becomes

\begin{align}
{\omega ^\mu }_{(T)} = (0,0,0,-\frac{{( - 1 + C)w}}
{{1 - {w^2}{r^2}}}).
\end{align}
When $C=0$ and torsion=0, the result is the vorticity in V4.
When $C=2$, the vorticity vector changes to the opposite sign.
\subsection{Other parts of torsion and spin-rotation coupling}
The traceless part of the torsion tensor can be further decomposed into the totally antisymmetric part (A) and the rest part (TL). The totally antisymmetric part (A) is given by
\begin{align}
{T^{(A)}}_{\mu \nu \sigma } = {T_{[\mu \nu \sigma ]}}
\end{align}
and the rest part (TL) is 
\begin{align}
{\tilde T_{\mu \nu }}^\lambda  = {T_{\mu \nu }}^\lambda  - \frac{2}
{3}\delta _{{\text{[}}\nu }^\lambda {T_{\mu {\text{]}}}} - \frac{1}
{3}{\varepsilon _{\mu \nu \beta }}^\lambda {\hat T^\beta }.
\end{align}
Calculating the vorticity vector in all the previous cases under other parts of torsion tensor, the results are listed in Table. II.\\
Table. II: The value of the torsion constant C which turns the
vorticity to the opposite sign.

{\raggedright
\vspace{3pt} \noindent
\begin{tabular}{p{39pt}p{39pt}p{39pt}p{39pt}p{39pt}p{35pt}}
\hline
\parbox{39pt}{\raggedright } & \parbox{39pt}{\raggedright 
Mink.
} & \parbox{39pt}{\raggedright 
Schwarz.
} & \parbox{39pt}{\raggedright 
Godel
} & \parbox{39pt}{\raggedright 
Kerr
} & \parbox{39pt}{\raggedright 
Lewis
} \\
\hline
\parbox{35pt}{\raggedright 
A+TL
} & \parbox{39pt}{\raggedright 
C=2
} & \parbox{39pt}{\raggedright 
C=2
} & \parbox{39pt}{\raggedright 

C=2
} & \parbox{39pt}{\raggedright 
C=2
} & \parbox{39pt}{\raggedright 
C=2
} \\
\parbox{39pt}{\raggedright 
A
} & \parbox{39pt}{\raggedright 
C=6
} & \parbox{39pt}{\raggedright 
C=6
} & \parbox{39pt}{\raggedright 
C=6
} & \parbox{39pt}{\raggedright 
C=6
} & \parbox{39pt}{\raggedright 
C=6
} \\
\parbox{39pt}{\raggedright 
TL
} & \parbox{39pt}{\raggedright 
C=3
} & \parbox{39pt}{\raggedright 
C=3
} & \parbox{39pt}{\raggedright 
C=3
} & \parbox{39pt}{\raggedright 
C=3
} & \parbox{39pt}{\raggedright 
C=3
} \\
\parbox{39pt}{\raggedright 
V
} & \parbox{39pt}{\raggedright 
X
} & \parbox{39pt}{\raggedright 
X
} & \parbox{39pt}{\raggedright 
X
} & \parbox{39pt}{\raggedright 
X
} & \parbox{39pt}{\raggedright 
X
} \\
\parbox{39pt}{\raggedright 
A+TL+V
} & \parbox{39pt}{\raggedright 
C=2
} & \parbox{39pt}{\raggedright 
C=2
} & \parbox{39pt}{\raggedright 
C=2
} & \parbox{39pt}{\raggedright 
C=2
} & \parbox{35pt}{\raggedright 
C=2
} \\
\hline
\end{tabular}
\vspace{2pt}

}

{\raggedright
{\footnotesize X: No Influence on vorticity, A: traceless part and rotationally antisymmetric, TL: traceless part but not totally antisymmetric, V: trace part.}
}
{\raggedright
{\footnotesize }
}

In the case of the traceless torsion (A + TL), (A), and (TL), the torsion constant $C$ = 2, 6, and 3 results in the
opposite vorticity, respectively. The trace part of the torsion will not influence the vorticity, so the full torsion has
the same constant $C$ = 2 as in the A + TL case which can induce minus vorticity.

In order to understand the deeper physical meaning of the opposite sign of the vorticity in U4, we calculate the axial torsion vector in all the above mentioned space-times;
they all have the same form
\begin{align}
{\hat T^\alpha } = \frac{C}
{3}{\omega ^\alpha } = \frac{C}
{3}{\Omega ^\alpha }_{(ob - gyro.)},
\end{align}
where $\Omega$ is the angular velocity, and (ob-gyro) means the quantity of the observer that relates to the gyroscope.
It should be mentioned that the spin density from Eqs.(12) and (15) is compatible with the torsion, Eq.(26), only in Godel space-time, and the proportionality between axial the torsion vector and the vorticity vector depends on the tetrad. For the Kerr metric, three common frames are: the locally non-rotating frame (LNRF), the Carter frame, and the local static frame [24]. The results of each frame are listed in Appendix A. Only the local static frame will satisfy Eq. (62).

It had been shown that the spin precession of a Dirac particle in the external torsion is \cite{K.Hayashi1979,J.Nitsch1980,Audretsch1981}
\begin{align}
\frac{{dS}}
{{dt}} =  - 3{\hat T^\alpha } \times S.
\end{align}
Substituting Eq.(62) into Eq.(63)
\begin{align}
\frac{{dS}}
{{dt}} =  - 3{\hat T^\alpha } \times S =  - C{\Omega _{(ob - gyro)}} \times S
\end{align}
which has the same form as the spin-rotation coupling \cite{Mashhoon1995,F.W.Hehl1990}. This effect describes the rotating observers with an angular velocity that would be associated with a Hamiltonian $H=-S \cdot \Omega$, where S is the intrinsic spin. Then the equation of motion $\hbar dS/dt = i[H,S]$ coincides with the precession equation of the classical spin vector $d{S_i}/dt =  - {\varepsilon _{ijk}}{\Omega _j}{S_k}$. Recently, it was shown \cite{Yu.N.Obukhov2009,Silenko2008} that the dynamics of classical and quantum spins in curved space-time are identical. It is frame independent within the order of Schiff's precession. According to our calculations, the presence of torsion will not change the magnitude of gyroscopic precession but will result in the opposite sign. It seems natural to interpret it as the precession of a quantum spin with positive and negative helicity. The inclusion of torsion in the classical treatment leads to a very similar behavior which is analogous to the property of quantum spin. This phenomenon was shown in another method by Pasini \cite{Pasini1990} who considered using an asymmetric connection.

Zhang and Beesham \cite{Zhang2002} recovered the spin-rotation coupling of the Dirac particle in the teleparallel theory. In
fact, it belongs to the case of $C$ = 1. When $C$ = 6, only the totally antisymmetric part of the torsion is involved; $dS/dt =  - 6\Omega  \times S$ should include the spin-rotation coupling and the effect between the Dirac particle and the torsion field. When C=2, the axial torsion still contributes to the spin-rotation coupling, but there should be other effects from the TL part of the torsion. On the other hand, the trace part of the torsion will not contribute to the opposite vorticity; this might resort to that it has no relation to spin. In addition, the helicity flip of the Dirac spin induced by the axial torsion was studied by
Capozziello $et$ $al$. \cite{1999}.

 It is important to note that the torsion effect has a different property between rotating and nonrotating space-time. The spin-rotation coupling appears in the rotating reference frame. For nonrotating space-time, the torsion effect which is in the shape of the spin-rotation coupling will vanish in the nonrotating frame. For rotating space-time, the angular velocity
${\Omega _{(ob - gyro)}} =  - {\Omega _{fg}}$ which is the angular velocity of the gyroscopic precession due to the frame dragging effect. According to the gravitational Larmor theorem \cite{B.Mashhoon1993}, the gyroscope undergoes a frame dragging precession $\Omega$ with respect to observers at infinity, the gravitomagnetic effects are locally equivalent to inertial effects in a frame rotating with frequency $-\Omega$. Thus, the torsion effect which is spin-rotation coupling shaped becomes
\begin{align}
dS/dt =  - C{\Omega _{(ob - gyro)}} \times S = C{\Omega _{fg}} \times S.
\end{align}
The effect will not vanish as long as the gravitomagnetic field, ${\text{H(u) = 2}}{\Omega _{fg}}$  \cite{D.Bini2005, F.D.Felice2010} exists.
It will lead to the presence of a spin-gravitomagnetism coupling term in the Hamiltonian and the gravitational Stern-Gerlach force which will be discussed in next section.

The unique property of torsion in rotating space-time could be seen alternately through the axial torsion vector.
The axial torsion vector is a measure of the deviation from spherical symmetry \cite{J.Nitsch1980}, but it depends on the
tetrad. For the Schwarzschild metric, the frame of the static observer reflects the spherical symmetry of spacetime
so the axial torsion vector is identical to zero. In our case, the axial torsion appears in the Schwarzschild metric due to the choice of the rotating reference frame. In contrast, for the Kerr metric, the axial torsion vector reflects the rotation of the space-time. The existence of the torsion effect is independent of the reference frame in rotating space-time.
\subsection{Gravitational Stern-Gerlach force}
Considering the spin-rotation coupling, the intrinsic spin $S$ of a particle would couple to the gravitomagnetic field of a rotating source via the interaction Hamiltonian
\begin{align}
H = S \cdot {\Omega _{fg}}.
\end{align}
It follows that the particle is subject to a gravitational Stern-Gerlach force given by
\begin{align}
F =  - \nabla (S \cdot {\Omega _{fg}}).
\end{align}
A spinning particle falls differently in the gravitational field of the Earth with spin-up and spin-down. In the
correspondence limit, this force can be deduced from the Mathisson-Papapetrou spin-curvature force
\cite{B.Mashhoon2001,Wald1972}. The motion of classical spinning particles in the gravitational field is described by the Mathisson-Papapetrou equations. There exists a spin-curvature force;
\begin{align}
{u^\nu }{\tilde \nabla _\nu }{P^\mu } =  - \frac{1}
{2}{\tilde R^\mu }_{\alpha \beta \gamma }{u^\alpha }{S^{\beta \gamma }} = {F^\mu }.
\end{align}
\begin{align}
{P^\mu } = m{u^\mu } - {u^\alpha }{\tilde \nabla _\alpha }{S^{\mu \nu }}{u_\nu },
\end{align}
where ${P^\mu }$ is the total four-momentum, m is the particles mass in the rest frame, and $S^{\mu\nu}$ is the antisymmetric classical spin tensor. It was shown \cite{A.O.Barut1987, Cianfrani2008} that the motion of the quantum spin particle coming out from the semiclassical limit of a spinor field coincides with a classical spinning particle at the lowest order in the spin. They all have the same spin-curvature coupling equation as Eq.(68). In the presence of torsion, this force becomes \cite{Hayashi1991}
\begin{align}
{F_{(T)}}^\mu  = {u^\nu }{\nabla _\nu }{P^\mu } =  - {T_{\alpha \beta }}^\mu {P^\alpha }{u^\beta } - \frac{1}
{2}{R^\mu }_{\alpha \beta \gamma }{u^\alpha }{S^{\beta \gamma }}.
\end{align}

There are several supplementary conditions to close the Mathisson-Papapetrou equations\cite{Filipe2002, DFilipe2012, K.Kyrian2007}. Three common choices are the Papapetrou-Corinadesi condition ${S^{t \nu}}=0$, the Pirani condition $S^{\mu\nu}u_{\nu}=0$, and the Tulczyjew condition $S^{\mu\nu}P_{\nu}=0$. Recently, the interpretations of the
Mathisson motions arising from the Pirani condition were given by Costa $et$ $al$  \cite{Filipe2002, DFilipe2012}. The origin of helical motions is the fact that the Pirani condition does not determine the worldline of the spinning particle uniquely. The radius of the helical motions is not arbitrarily large and should be contained within the minimum size of the classical spinning particle.    

First, we analyze the spin-curvature force of the Godel space-time in V4. The components of spin-curvature force are
\begin{align}
{F^\mu } = (\frac{{-{e^x}{S^{ty}}}}
{{2a}},\frac{{ {S^{tx}} - {e^x}{S^{xy}}}}
{{2a}},\frac{{ {S^{ty}}}}
{{2a}},0).
\end{align}
The left-hand side of Eq.(68) is given by
\begin{align}
\begin{gathered}
  {P^\mu } =  \hfill \\
  t:(  m + a{e^x}{{\dot S}^{ty}} - a{S^{tx}} + a{e^x}{S^{xy}})/a \hfill \\
  x: - {{\dot S}^{tx}} + {e^x}{{\dot S}^{xy}} - {e^x}{S^{ty}}/2 \hfill \\
  y:( - {e^x}{{\dot S}^{ty}} + {S^{tx}} - {e^x}{S^{xy}})/{e^x} \hfill \\
  z: - {{\dot S}^{tz}} + {e^x}{{\dot S}^{zy}} \hfill, \\ 
\end{gathered}
\end{align}
\begin{align}
\begin{gathered}
  {u^\alpha }{{\tilde \nabla }_\alpha }{P^\mu } =  \hfill \\
  t:-(-2\dot m + 4a{{\dot S}^{tx}} - 2a{e^x}{{\ddot S}^{ty}} - 4a{e^x}{{\dot S}^{xy}} + a{e^x}{S^{ty}})/2{a^2} \hfill \\
  x:( - 2{{\ddot S}^{tx}} - 2{e^x}{{\dot S}^{ty}} + 2{e^x}{{\ddot S}^{xy}} + {S^{tx}} - {e^x}{S^{xy}})/2a \hfill \\
  y:-(2{e^x}{{\ddot S}^{ty}} - 4{{\dot S}^{tx}} + 4{e^x}{{\dot S}^{xy}} - {e^x}{S^{ty}})/2a{e^x} \hfill \\
  z:-({{\ddot S}^{tz}} - {e^x}{{\ddot S}^{zy}})/a \hfill \\ 
\end{gathered}
\end{align}
where a dot denotes differentiation with respect to t.
By equating the left-hand side to the right-hand side of Eq. (68), four simultaneous equations are obtained
\begin{align}
\left\{ {\begin{array}{*{20}{c}}
   { - 2{e^x}{{\ddot S}^{ty}} + 4{{\dot S}^{tx}} - 4{e^x}{{\dot S}^{xy}} = 0}  \\
   { - 2{e^x}{{\dot S}^{ty}} - 2{{\ddot S}^{tx}} + 2{e^x}{{\ddot S}^{xy}} = 0}  \\
   {{{\ddot S}^{tz}} - {e^x}{{\ddot S}^{zy}} = 0}  \\
   {\dot m = 0}  \\
 \end{array}}.
\right.\end{align}
Using the Pirani supplementary condition 
\begin{align}
S^{\mu \nu }{u_\nu } = 0,
\end{align}
which implies
\begin{align}
\dot m = 0,{S^{ty}} = 0.
\end{align}
In Godel cosmology, it is reasonable to set $\dot x = 0$ \cite{Smalley1985}.
In that case, the constrained conditions Eq.(74) reduce to
\begin{align}
\left\{ {\begin{array}{*{20}{c}}
   {{{\dot S}^{tx}} - {e^x}{{\dot S}^{xy}} = 0}  \\
   {{{\ddot S}^{tz}} - {e^x}{{\ddot S}^{zy}} = 0}  \\
\end{array} }.
\right.\end{align}
Now we consider full torsion in U4, the spin-curvature force becomes
\begin{align}
\begin{gathered}
  {F_{(T)}}^t = \frac{{{{-( - 1 + C)}^2}{e^x}{S^{ty}}}}
{{2a}} \hfill, \\
  {F_{(T)}}^x = \frac{{{{( - 1 + C)}^2}(  {S^{tx}} - {e^x}{S^{xy}})}}
{{2a}} \hfill, \\
  {F_{(T)}}^y = \frac{{  {{( - 1 + C)}^2}{S^{ty}}}}
{{2a}},{F_{(T)}}^z = 0 \hfill. \\ 
\end{gathered}
\end{align}
For $C$=0 and $C$=2, the force will not change sign. This is because the torsion operates twice on the left-hand side. 
The four-momentum in U4 changes to
\begin{align}
\begin{gathered}
  {P_{(T)}}^\mu  =  \hfill \\
  t:(  m + a{e^x}{{\dot S}^{ty}} + ( - 1 + C)a{S^{tx}} - ( - 1 + C)a{e^x}{S^{xy}})/a \hfill \\
  x: - {{\dot S}^{tx}} + {e^x}{{\dot S}^{xy}} + ( - 1 + C){e^x}{S^{ty}}/2 \hfill \\
  y:( - {e^x}{{\dot S}^{ty}} - ( - 1 + C)({S^{tx}} - {e^x}{S^{xy}}))/{e^x} \hfill \\
  z: - {{\dot S}^{tz}} + {e^x}{{\dot S}^{zy}} \hfill. \\ 
\end{gathered}
\end{align}
Using the conditions, Eqs.(76)-(78), for $C$=2, the following result is derived:
\begin{align}
\left\{ {\begin{array}{*{20}{c}}
   {{u^\alpha }{{\tilde \nabla }_\alpha }{P^\mu } = {F^\mu }}  \\
   {{u^\alpha }{\nabla _\alpha }{P^\mu } =  - {F^\mu }}  \\
\end{array} }. 
\right.\end{align}

In quantum physics, the opposite Stern-Gerlach force is related to the spin with the opposite helicity. It provides a
good analogue to interpret the opposite sign of the spincurvature force as the behavior of a different helicity of
spin which supports our suggestion about the opposite gyroscopic precession.
\subsection{ECSK equations in vacuum solution}
In the vacuum solution in V4, for example, the Schwarzschild and Kerr metrics, their Einstein equations should take the form
\begin{align}
{\tilde G_{\mu \nu }} = 0.
\end{align}
The Einstein tensor and Ricci curvature are zero due to the vacuum, but the fourth rank Riemann curvature tensor is not zero because the curvature will propagate.
The Einstein equation of vacuum solution in U4 becomes

\begin{align}
{\tilde G_{\mu \nu }} + Torsion ~term  = {G_{\mu \nu }} = k({T_{\mu \nu }} + {T_{\mu \nu (spin)}}),
\end{align}
where ${\tilde G_{\mu \nu }}$  is zero due to vacuum which is ${T_{\mu \nu }} = 0$. If ${G_{\mu \nu }}=0$, the only possible solution is ${T_{\mu \nu (spin)}}=0$. All the matters which can induce curvature and torsion are zero. This is the vacuum in U4.

First, we consider the traceless part of torsion (A + TL). The nonvanishing components of the Einstein tensor of the
Schwarzschild metric in U4 are listed in Appendix B. In Eq. (B1), all components are proportional to torsion constant
$C$; when $C$ = 0, it will reduce to the vacuum solution in V4, thus the Einstein tensor is zero. Now we turn to calculate the case with full torsion (A + TL + V) in U4, The nonzero components of the Einstein tensor in U4 with full torsion are
\begin{align}
\begin{gathered}
  {G_{tt}} = \frac{{C( - 1 + C){{( - r + 2m)}^2}}}
{{{r^4}}}, \hfill \\
  {G_{t\phi }} = \frac{{ - C( - 1 + C)( - r + 3m)mw{{\sin }^2}\theta }}
{{r(r - 2m - {w^2}{r^3}{{\sin }^2}\theta )}}, \hfill \\
  {G_{rr}} = \frac{{C( - 1 + C)}}
{{r( - r + 2m)}}, \hfill \\
  {G_{\theta r}} = -\frac{{C( - 1 + C)\cos \theta }}
{{r\sin \theta }},{G_{\theta \theta }} = -\frac{{C( - 1 + C)m}}
{r}, \hfill \\
  {G_{\phi t}} = \frac{{C( - 1 + C)w( - r + 2m)(r - 3m{{\sin }^2}\theta )}}
{{r(r - 2m - {w^2}{r^3}{{\sin }^2}\theta )}}, \hfill \\
  {G_{\phi \phi }} =- \frac{{C( - 1 + C)m{{\sin }^2}\theta }}
{r}. \hfill \\ 
\end{gathered}
\end{align}
When $C$ = 2, it will cause the vorticity to the opposite sign which could be interpreted as the spin flip; thus, the spacetime
might have particle spin prior and induce space-time torsion. When $C$ = 0, torsion = 0, and all the components of the Einstein tensor are zero, we get the vacuum of the Schwarzschild metric in V4 that is ${\tilde G_{\mu \nu }}=0$. However, there is still another condition to make the vacuum in U4. For $C$ = 1, the Einstein tensor also vanishes, but the torsion tensor still exists. As we discussed before, the vanishing Einstein tensor of the vacuum solution in U4 arises from ${T_{\mu \nu (spin)}}=0$.It implies that the full torsion will propagate, but the traceless part of the torsion will not. The result actually obeys the principle of the contact interaction of spin \cite{Hehl1976}in ESCK theory and still resorts to the reason that only the traceless torsion couples to spin. The same result can also be obtained in the Kerr metric,which is the exterior vacuum solution of the rotating body.

Using Eq.(20), it has the same form as that in U4 with respect to U4 connection \cite{Palle1999}:
\begin{align}
{\nabla _\alpha }{\omega ^\alpha }_{(T)} = 2{a_\alpha }_{(T)}{\omega ^\alpha }_{(T)}.
\end{align}
In the case of the traceless part of the torsion, when $C$=2, ${\omega ^\alpha }_{(T)} =  - {\omega ^\alpha }$ then
\begin{align}
- {\tilde \nabla _\alpha }{\omega ^\alpha } =  - 2{a_\alpha }_{(T)}{\omega ^\alpha }
\end{align}
thus
\begin{align}
{a_\alpha }_{(T)} = {a_\alpha }.
\end{align}
If the observer moves from V4 to U4 space-time (A+TL torsion), it will feel the same 
acceleration but has a chance (very low probability in a weak field) to observe the helicity flip of a comoving spin. 

Now we turn to consider the situation with the full torsion, for $C$=2
\begin{align}
{\nabla _\alpha }{\omega ^\alpha }_{(T)} =  - {\tilde \nabla _\alpha }{\omega ^\alpha } + {K_{\alpha \beta }}^\alpha {\omega ^\beta } =  - 2{a_{(T)}}{\omega ^\alpha }
\end{align}
thus
\begin{align}
2({a_\alpha }_{(T)} - {a_\alpha }){\omega ^\alpha } =  - {K_{\alpha \beta }}^\alpha {\omega ^\beta }.
\end{align}
The trace part of the torsion contributes to the difference of acceleration that the observer feels between V4 and U4.
\section{Discussions and Conclusions}
In the Godel space-time, the vorticity vector is the global rotation, which means the whole Galaxy with spinning fluid rotates with the same angular velocity. After calculating the vorticity vectors of specific observers in other rotating space-times, we eventually realize the angular velocity which can flip to the opposite sign is the gyroscopic precession caused by the frame dragging effect. In nonrotating space-time, the Schwarzschild and Minkowski metrics, the minus sign will also couple to the de Sitter and Thomas precessions.

Smalley attributed the minus vorticity to the frame dragging effect that appears in the static space-time by adding spinning fluid \cite{Smalley1985}. If we image the minus vorticity as an opposite gyroscopic precession, for the rotatingmetric,
it means the opposite gravitomagnetic field. It seems unreasonable when the rotating gravitational source is not
changed. We suggest that the better interpretation is the dynamics with respect to the helicity of spin. The opposite
gyroscopic precession could be imaged as the classical analogy of the precession of the particle with positive and
negative helicity.

Analyzing the Mathisson-Papapetrou equations of the Godel space-time, the torsion constant which can induce
minus vorticity will also induce a minus spin-curvature force. The spin-curvature force is the gravitational Stern-Gerlach force in the weak field limit. It is natureal to link the force with an opposite direction to the behavior of particles with spin-up and -down which is analogous to the Stern-Gerlach force of the quantum spin. It supports our previous suggestion and is also consistent with the results that the dynamics of classical and quantum spins in curved space-time are identical\cite{Yu.N.Obukhov2009,Silenko2008}. In our purely classical treatment, the inclusion of torsion tensor results
with a very similar property of quantum spin should resort to ECSK theory, which states that spin is the source of the
torsion.

Although we did not assume the spin fluid energy momentum tensor prior like other models \cite{Smalley1985,Tsoubelies1984}, the nonzero torsion might imply that intrinsic spins have already existed. The trace part of the torsion cannot contribute to the opposite sign of the vorticity because it has no relation with the particle spin. The whole traceless part will induce
helicity flip; it includes very similar results of spin-rotation coupling caused by totally antisymmetric torsion and other
contributions from the rest part of the traceless torsion. In nonrotating space-time, the torsion effect in the shape of
the spin-rotation coupling will vanish if the reference frame is nonrotating. In the rotating space-time, the torsion
effect in the shape of the spin-gravitomagnetism coupling was obtained through the local static observer. This term
will not vanish as long as the gravitomagnetic field exits in in the space-time. It might imply the result of some authors \cite{C.M.Zhang1992,Kuchowicz1976}that the torsion also originates from macroscopic rotation, and we suggest the torsion effect is similar to the coupling of spin and macroscopic rotation. Moreover, we show that the full torsion will propagate in vacuum. It is demonstrated that the torsion trace part plays the essential role of propagating torsion and the traceless part contributes
to spin-spin contact interaction.

\appendix
\section{Axial torsion and vorticity vector for Kerr metric with different frames}
Table. I: The axial torsion and vorticity vector for Kerr metric with different frames.

{\raggedright
\vspace{3pt} \noindent
\begin{tabular}{p{34pt}p{100pt}p{100pt}}
\hline
\parbox{39pt}{\raggedright } & \parbox{100pt}{\raggedright 
Axial torsion vector
} & \parbox{100pt}{\raggedright 
Vorticity vector
} \\
\hline
\parbox{34pt}{\raggedright 
LNRF
} & \parbox{100pt}{\raggedright 
${\hat T}^r= 2Cmr{a^3}cos\theta si{n^2}\theta \Delta$ \\
$ /3{\Sigma ^2}({r^4} + {r^2}{a^2} + 2mr{a^2} + {a^2}co{s^2}\theta \Delta )$
${{\hat T}^\theta } =  - Cmasin\theta ({a^4}co{s^2}\theta  - {a^2}{r^2}co{s^2}\theta  - {a^2}{r^2} - 3{r^4})
  /3{\Sigma ^2}({r^4} + {r^2}{a^2} + 2mr{a^2} + {a^2}co{s^2}\theta \Delta )$
} & \parbox{100pt}{\raggedright 
0
} \\
\parbox{39pt}{\raggedright 

} & \parbox{39pt}{\raggedright 

} & \parbox{39pt}{\raggedright 

} \\
\parbox{39pt}{\raggedright 
Carter
} & \parbox{100pt}{\raggedright 
${\hat T^r} = Ca\Delta \cos \theta /3{\Sigma ^2}$
${\hat T^\theta } = Car\sin \theta /3{\Sigma ^2}$
} & \parbox{100pt}{\raggedright 
${\omega ^r} = a\Delta \cos \theta /{\Sigma ^2}$
${\omega ^\theta} =0$} \\
\parbox{39pt}{\raggedright 

} & \parbox{39pt}{\raggedright 

} & \parbox{39pt}{\raggedright 

} \\
\parbox{39pt}{\raggedright 
Local static frame
} & \parbox{39pt}{\raggedright 
${\hat T^r} =  - 2Cmar\cos \theta \Delta/3{\Sigma ^2}(\Sigma-2mr)$
${\hat T^\theta } = -Cma\sin \theta({r ^2}-{a^2}co{s^2}\theta )/3{\Sigma ^2}(\Sigma  - 2mr)$
} & \parbox{39pt}{\raggedright 
see Eq.(35)
} \\
\hline
\end{tabular}
\vspace{2pt}
{\raggedright
{\footnotesize LNRF: locally non-rotating frame, Carter: the Carter frame}
}
{\raggedright
{\footnotesize }
}

\section{The Einstein tensor of Schwarzchild metric with traceless torsion}
We define the notation  ${\Gamma}=r-2m$.
\begin{align}
\begin{gathered}
  {G_{tt}} = C(( - 2Cr{\Gamma}  + (3C{r^2} - 10Crm + 9C{m^2}\hfill \\- 3{r^2}
   - 18{m^2} + 12rm )si{n^2}\theta))/9{r^4}si{n^2}\theta\hfill \\
  {G_{\phi t}} =- C(((- 2C+3)wr{\Gamma}  
  + (2Cwrm- 9Cwrm+ Cw{r^2}\hfill \\ - 9Cw{m^2} + 18w{m^2} )si{n^2}\theta ))
  /( - 3r{\Gamma} + 3{w^2}{r^4}si{n^2}\theta ) \hfill \\
  {G_{rr}} = -C((  2Cr{\Gamma} + (C{r^2} - 14Crm + 27C{m^2}+ 42rm\hfill \\ - 54{m^2}
   - 9{r^2} )si{n^2}\theta ))/( 36{r^3}m - 36{r^2}{m^2}- 9{r^4} )si{n^2}\theta \hfill \\
  {G_{\theta r}} =   Ccos\theta (3Cm- Cr +3{\Gamma} )/9r{\Gamma}sin\theta \hfill \\
  {G_{r\theta }} =   Ccos\theta (3Cm- Cr - 6{\Gamma} )/9r{\Gamma}sin\theta \hfill \\
  {G_{\theta \theta }} =C((  6Crm-3C{r^2}+6r{\Gamma} + ( 6Crm - 18C{m^2}+ C{r^2}\hfill \\ -6{r^2} + 21rm - 18{m^2})si{n^2}\theta ))/ - 9r{\Gamma}si{n^2}\theta \hfill \\
  {G_{t\phi }} = -C((-Cwr{\Gamma}  + (10Cwrm - Cw{r^2} - 18Cw{m^2}\hfill \\ - 3wrm
   + 9w{m^2})si{n^2}\theta ))/( -3r{\Gamma} + 3{w^2}{r^4}si{n^2}\theta ) \hfill \\
  {G_{\phi \phi }} =-C((-C+6)r{\Gamma} + ( - 14Crm+ 3C{r^2} + 18C{m^2} \hfill \\- 9m{\Gamma}
   )si{n^2}\theta )/-9r{\Gamma} \hfill. \\ 
\end{gathered}
\end{align}

\section*{Acknowledgements}
We would like to thank Alexander Silenko, Donato Bini, and Jun Chu Chen for helpful discussions. We would also like to thank  Jiun-Huei Proty Wu and Pisin Chen for providing many useful suggestions to improve our paper.

\end{document}